\documentstyle[12pt,twoside]{article}
\def\beq{\begin{equation}}
\def\eeq{\end{equation}}
\def\bear{\begin{eqnarray}}

\def\bearr{\begin{eqnarray} &&}
\def\ear{\end{eqnarray}}

\def\dst{\displaystyle}
\def\chih{{\chi\mathstrut}}
\def\nn{\nonumber\\ }
\def\nnn{\nonumber\\ && {} }
\newcommand{\vars}[1]{\left\{\begin{array}{ll}#1\end{array}\right.}

\def\sign{{\,\rm sign}\,}
\def\const{{\rm const}}

\def\dim{{\,\rm dim\,}}
\def\arccot{{\,\rm arccot\,}}
\def\e{{\,\rm e}}
\def\to{\rightarrow}

\def\lst{\lambda_{\rm string}}

\def\sumi{\sum_{i=1}^{n}}
\def\half{{\textstyle\frac{1}{2}}}
\def\g{\hat g}
\def\ve{\varepsilon}
\def\hx{\hat \xi}
\def\ds{d{\hat s}^2_D}
\def\go{\overline{g}}
\def\Ro{\overline{R}}
\def\dso{d\overline{s}^2}
\def\axi{axially symmetric\ }
\def\sph{spherically symmetric\ }
\def\wh{wormhole}
\def\sqat{\sqrt{a_3}}

\begin{document}
\begin{center}

{\Large\bf RING WORMHOLES IN D-DIMENSIONAL \\[5pt]
           EINSTEIN AND DILATON GRAVITY}
\medskip

{\bf Kirill A. Bronnikov\footnote{Permanent address:
Center for Gravitation and Fundamental Metrology, VNIIMS,
     3--1 M. Ulyanovoy Str., Moscow 117313, Russia;
	e-mail: bron@cvsi.rc.ac.ru}
and J\'ulio C. Fabris}
\medskip

Departamento de F\'{\i}sica, Universidade Federal do Esp\'{\i}rito Santo, 
Vit\'oria, ES,
Brazil\footnote{\e-mail: fabris@npd2.ufes.br}

\end{center}

\begin{abstract}
     On the basis of exact solutions to the
     Einstein---Abelian gauge---dilaton equations in $D$-dimensional gravity,
     the properties of static  \axi  con\-fi\-gu\-ra\-ti\-ons are discussed.
     Solutions free of curvature singularities are
     selected; they can be attributed to traversible wormholes with
     cosmic string-like singularities at their necks.  In the
     presence of an electromagnetic field some of these wormholes are
	globally regular, the string-like singularity being replaced by a set
	of twofold branching points. Consequences of
     wormhole regularity and symmetry conditions are discussed.
	In particular, it is shown that (i) regular, symmetric \wh s have
     necessarily positive masses as viewed from both asymptotics and (ii)
	their characteristic length scale in the big charge limit
	($GM^2 \ll Q^2$) is of the order of the ``classical radius" $Q^2/M$.
\end{abstract}

\section{Introduction}


Wormholes as regular 3-geometries with a neck connecting two flat asymptotics
have apparently first attracted the researchers' attention as particular
spatial sections of the extended Schwarzschild (Kruskal) space-time. The
possible existence of such geometries invoked such attractive ideas as
``charge without charge'' or ``mass without mass'' \cite{wh}. However,
the Schwarzschild and Reissner-Nordstrom wormholes are not traversible 
for a
non-tachyonic particle due to the dynamic nature of the relevant space-time.
Later on appeared static, \sph \wh\ metrics decscribing traversible \wh s
\cite{br-acta,hell}, but at the expense of introducing unusual types of
matter (scalar) fields, since in a certain domain near a \wh\ neck one must
have $\ve + p_r <0$ where $\ve$ and $p_r$ are the energy density and
pressure in the radial direction, respectively. Other \sph \wh s
were found in Refs. \cite{br-ann79} (in particular, a model with
electric field and a domain of neutral dust with negative density),
\cite{bryba} (as particle-like solutions of a class of
models of nonlinear electrodynamics), \cite{cle} (in higher-dimensional
models with sigma-model-type behavior of effective scalar fields in a
4-dimensional reformulation of the theory), \cite{br95a} (in
multidimensional dilaton gravity) and some others. Clement \cite{cle}
also obtained an axially symmetric generalization of his solutions,
describing a chain of \wh s situated along a symmetry axis.

In the recent years \wh s have been discussed as possible time machines
(\cite{machine} et al.); the discussion concerned the consequences
of \wh\ existence rather than models where they can appear.
This interesting set of problems is beyond the scope of this paper.
We also do not touch upon the vast recent work concerning Euclidean and
quantum wormholes.

Instead, we would like to discuss a class of Lorentzian, \axi \wh s having
no \sph analog --- we suggest to call them {\sl ring wormholes\/} --- whose
distinctive feature is the existence of a ring of branching points
like those in Riemannian surfaces of analytic functions. Unlike \sph
counterparts, these ones do not require unusual matter for their existence:
in the field model under consideration a necessary condition for their
regularity is just a kind of equilibrium condition between the electric and
scalar fields.

The field model, used here in a phenomenological manner, is
$D$-dimensional dilaton gravity. The latter (more precisely,
Einstein-gauge-dilaton-axion gravity) is known to form the bosonic part
of effective low energy string theory \cite{green}, one of the theories
pretending to become ``a theory of everything''. This is the main
motivation for a large number of studies of its solutions and
predictions (see, for instance, \cite{ksh2,galz} and references
therein).
     Another problem of interest (in string theory among others)
     is that of possible effects and manifestations of extra space-time
     dimensions. One of ways to take them into acount is to treat
     extra-dimension scale factors as separate dynamic variables,
     as is done, e.g., in
\cite{br-ann,br-vuz91,br10,br95a,br95b,shira,cadoni}.
	This is the approach adopted here.

     We start from the action
\beq                                                     
     S= \int d^D x \sqrt{^D g}\Big[{}^DR+g^{MN}\varphi_{,M}\varphi_{,N}
               -\e^{2\lambda \varphi}F^2\Big]              \label{Action}
\eeq
     where $g_{MN}$ is the $D$-dimensional
     metric, $^D g=\bigl|\det g_{MN}\bigr|$,\ $\varphi$ is the dilaton scalar
     field and $F^2 =  F^{MN}F_{MN},\
     F= dU$ is an Abelian gauge field,
     to be interpreted as the electromagnetic field.

     The field-theoretic limit of string theory corresponds to the specific
     value of the coupling constant $\lambda = \lst = \pm (D-2)^{-1/2}$
     \cite{green,shira}. However, we retain an arbitrary
     value of  $\lambda$ in order to cover a wider spectrum of possible field
     theories, such as Kaluza-Klein type ones, considered, e.g., in
     Ref.\,\cite{vladim}. The value $\lambda =0$ evidently corresponds to
	$D$-dimensional (in particular, 4-dimensional) general relativity (GR)
	with a minimally coupled scalar field.

	It should be noted that the action (\ref{Action}) is written in the
	so-called Einstein conformal gauge, convenient for solving the field
     equations. However, if the underlying
	theory is string theory, then a more fundamental role is played 
by the
	``string metric'', or ``$\sigma$ model metric''
\beq           \label{MapS}                                    
	\g_{AB}=\e^{-2\lambda\varphi}g_{AB}
\eeq
     rather than $g_{AB}$ from \ref{Action} (see, e.g., 
	\cite{shira,banks}
	and references therein). Therefore it makes sense to discuss
     such conformal gauge-dependent issues as the nature of singularities
	(if any) in terms of $\g_{AB}$. Strictly
	speaking, this argument applies only to $\lambda = \lst$,
     but, for convenience, we deal with $\g_{AB}$  for any
     $\lambda$, keeping in mind that this automatically leads to correct
	results for the case of GR ($\lambda =0$) as well.

     Throughout the paper capital Latin indices range from 0 to $D-1$, Greek
     ones from 0 to 3.

\section{Static, \axi solutions to the field equations} 
\subsection{Equations}                     

     Consider a $D$-dimensional pseudo-Riemannian manifold
     $V^D$ with the structure
\beq                                             
     V^D = M^4 \times V_1\times \ldots
     \times V_n; \ \ \dim V_i=N_i; \ \ D=4 + \sumi N_i,   \label{Stru}
\eeq
     where $M^4$ plays the role of the conventional
     space-time and $V_i$ are Ricci-flat manifolds of arbitrary dimensions
     and signatures, with the line elements $ds_i^2$, $i=1,\ldots,n$.
     The $D$-metric is
\beq                                             
     ds_D^2 = g_{MN}dx^M dx^N=
     g_{\mu\nu}dx^\mu dx^{\nu}
          + \sumi \e^{2\beta_i(x^\mu)}ds_i^2.         \label{DsD}
\eeq

     For studying static, \axi systems let us use the 4-dimensional
	formulation of the theory (\ref{Action}). If the gauge
     field $F$ is purely 4-dimensional (only $F_{\mu\nu}\ne 0$), the action
     (\ref{Action}) after integrating out the extra-dimension coordinates
	reads (up to a constant factor and a divergence):
\beq                                                     
     S=\int d^4 x \sqrt{^4 g}
     \e^{\sigma}\Bigl({}^4 R-\sigma^{,\mu}\sigma_{,\mu}
     +\sumi N_i\beta_{i,\mu}\beta_i{}^{,\mu} + \varphi^{,\mu}\varphi_{,\mu}
          -\e^{2\lambda\varphi} F^2 \Bigr)    \label{Act4}
\eeq
     where $^4 R$ is the
     curvature derived from $g_{\mu\nu}$, the 4-dimensional part of $g_{MN}$,
     and, as before, $\sigma= \sumi N_i\beta_i$.
     Eq.\,(\ref{Act4}) corresponds to the original $D$-Einstein
     conformal gauge. The 4-dimensional Einstein gauge with the metric
     $\go_{\mu\nu}$ is obtained after the conformal mapping
\beq                                                      
     \go_{\mu\nu}= \e^{\sigma}g_{\mu\nu}                      \label{Map4E}
\eeq
     leading the action to the form
\beq                                                           
     S=\int d^4 x \sqrt{{}^4 \go}
     \Bigl({}^4 \Ro + \half\sigma^{,\mu}\sigma_{,\mu}
     +\sumi N_i\beta_{i,\mu}\beta_i{}^{,\mu}	+ \varphi^{,\mu}\varphi_{,\mu}
          -\e^{\sigma+2\lambda\varphi} F^2 \Bigr)             \label{Act4E}
\eeq
     where $\go_{\mu\nu}$ is used to form the curvature ${}^4\Ro$ and to
	raise and lower the indices.

	Let us now adopt the following assumptions:
\begin{description}
\item[(i)] the static, \axi
	metric $\go_{\mu\nu}$ takes the Weyl canonical form
\beq                                                                
	\dso = \e^{2\gamma}dt^2-\e^{-2\gamma}
	[\e^{2\beta}(d\rho^2 + dz^2) + \rho^2 d\phi^2];   \label{DsA}
\eeq
\item[(ii)] the gauge field $U$ has the only component $U_0 = U(\rho,z)$
	(that is, only the electric field is present);
\item[(iii)] there is only one ``extra'' space $V_1$ (we will denote $\dim
	V_1= d,\ \beta_1=\xi$).
\end{description}
	Assumption (ii) is not very restrictive since a magnetic
	component of $F$, compatible with axial symmetry and
	regularity on the axis, may be easily introduced by a duality rotation
	applied to the purely electric field $F$.
	Assumption (iii) is adopted mostly to save space, since, as seen
	from (\ref{Act4E}), a generalization to $n > 1$ is straightforward.

	The vacuum field equations (for $\varphi\equiv U \equiv 0$) can be
	written in the form \cite{br95b}
\bear                                                     
	\Delta \gamma &=& \Delta \xi =0;    \nn
     \beta_\rho  &=& \rho [\Delta_1 \gamma
		+ \half a_0 \Delta_1 \xi],
          \quad \Delta_1 \gamma \equiv \gamma^2_\rho - \gamma^2_z,
                       \ \ {\rm etc.}, \nn
	\beta_z &=& [2\gamma_\rho \gamma_z + a_0 \xi_\rho \xi_z] \label{Evac}
\ear
	where the indices $\rho$ and $z$ denote partial derivatives, $a_0 =
	\half d^2 + d$ and $\Delta$ is the ``flat'' Laplace operator in the
	cylindrical coordinates:
\[
    \Delta = \rho^{-1}\partial_\rho(\rho\partial_\rho)+\partial_z\partial_z.
\]
	Eqs.\,(\ref{Evac}) coincide with the scalar-vacuum equations in
	conventional GR; the ``scalar field'' $\xi$
	enters into the equations in the same way as the metric function
	$\gamma$, and, just as in GR, their solution by
	quadratures reduces to that of the flat-space Laplace	equation
	\cite{Synge}.

	It can be shown that
	in the general case (\ref{Act4E}), under a certain additional
	assumption, the field equations essentially reduce to (\ref{Evac}).
        Indeed, introduce the combinations of the unknowns $\gamma$,
	$\xi$ and $\varphi$
\bear
	\chi  &=& -(d+2)\lambda\xi + \varphi, \nn
     \nu &=& \gamma + {\dst\frac{d+2}{2a_1}}(d\xi + 2\lambda\varphi),\nn
	\omega &=& \gamma - \half d\xi -\lambda\varphi,        \label{Tuda}
\ear
	and assume that\footnote{As will be seen further,
        $\e^{2\omega(\rho,z)}=\g_{00}$ is the
	static gravitational potential in string metric. Therefore
	Assumption (iv) means that the gravitational and electromagnetic field
	strength vectors are everywhere parallel. In particular, these two
	fields must have common sources, if any.}
\begin{description}
\item[(iv)] the potential $U(\rho,z)$ is functionally related to $\omega$.
\end{description}

	Then the field equation for $U$ takes the form of a linear first-order
	equation with respect to $f(\omega) = (dU/d\omega)^2$, which gives
\beq                                    %
	\e^{2\omega(\rho,z)} = a_3(c_0 + 2c_1 U + U^2)      \label{Om}
\eeq
	where $c_0$ and $c_1$ are integration constants.
	Defining the function $W(U)$ by the relation
\beq                               %
	dW/dU = \e^{-2\omega},                              \label{DefW}
\eeq
	it is an easy matter to bring the remaining field equations
	to a form like (\ref{Evac}):
\bear                                                \label{Edil}
	\Delta \chi &=& \Delta \nu = \Delta W =0;                     \nn
     \beta_\rho &=& \rho\Bigl[ {\dst\frac{d}{2a_1}}\Delta_1 \chih
          + {\dst\frac{1}{a_2}}\Delta_1\nu + K_1 \Delta_1 W\Bigr], \nn
     \beta_z &=& \rho\Bigl[{\dst \frac{d}{a_1}}\chih_\rho\chih_z
          + {\dst \frac{2}{a_2}} \nu_\rho \nu_z +2K_1 W_\rho W_z\Bigr]
\ear
	where we have introduced the constants
\bear
	a_1 &=&  d + 2\lambda^2 (d+2);\qquad	a_2 = 1 + 
\frac{d+2}{a_1}; \nn
	a_3 &=&  1+ \frac{a_1}{d+2}=\frac{a_1a_2}{d+2};
     \qquad K_1 = a_3(c_1^2 -c_0).                       \label{DefAK}
\ear
	A similar procedure was applied to this field system in
	4-dimensions in Ref.\,\cite{rash}.
	As in (\ref{Evac}), the integrability conditions for $\beta$ in
	(\ref{Edil}) are fulfilled automatically.
	Thus the solutions are obtained by quadratures
     provided the harmonic functions $\chih,\ \nu,\ W$ are known.
	The original functions $\gamma$, $\xi$ and $\varphi$ are restored
	using the reverse transformation
\bearr
	\gamma = \nu/a_2 + \omega/a_3,                       \nnn
	\xi    = -2\lambda\chih/a_1 + 2(\nu-\omega)/(a_1a_2),\nnn
	\varphi= d\cdot\chi/a_1+ 2\lambda(\nu-\omega)/a_3 \label{Obratno}
\ear
	(note that the determinant of (\ref{Tuda}) is $a_1a_2=a_1+d+2 >0$).

It should be noted that in the special case of our field model
$d = \lambda = 0$ (GR, $D=4$, scalar electrovacuum), when the above
formulas are not directly applicable due to $a_1=0$, Eqs.\,(\ref{Edil})
still hold with $d/a_1 =1$, $\omega \equiv \gamma$ and
$\chi \equiv \varphi$, while the coefficient $1/a_2$ vanishes and the
function $\nu$ does not appear at all. So all the subsequent formulas
are also valid with these substitutions; in particular, in the monopole
solutions of Subsec.\,2.3 one should put $q_2=0$ since this is the
``charge" corresponding to $\nu$.

\subsection{Multipole solutions}  

	Following \cite{axial,Rad77}, let us seek solutions in the
	coordinates $(x,y)$, connected with $\rho$ and $z$ by
\beq
     \rho^2 = L^2(x^2 +\varepsilon)(1-y^2),  \qquad
     z = Lxy                                          \label{XY}
\eeq
	where $L$ is a fixed positive constant and $\varepsilon= 0,\ \pm 1$,
	so that $x$ and $y$ are spherical ($\varepsilon=0$), prolate
     spheroidal ($\varepsilon=-1$), or oblate spheroidal ($\varepsilon=+1$)
     coordinates, respectively; $-1 < y < 1$; $x>1$ for $\varepsilon=-1$ and
	$x>0$ for $\varepsilon= 0,\ +1$.

	The Laplace equation $\Delta f =0$ acquires the form
\beq
     \partial_x(x^2+\varepsilon)\partial_x f
             +\partial_y(1-y^2)\partial_y f =0.          \label{Delta}
\eeq
     Separating the variables in Eq.(\ref{Edil}) (the first line), i.e.,
     putting, for example, $\chi (x,y) = \chih_1 (x)\chih_2 (y)$, one obtains
\bear                              
      {[(x^2 +
     \varepsilon)\chih_{1x}]}_x - \lambda_0\chih_1 &=& 0,  \label{Ex} \\
     {[(1-y^2)\chih_{2y}]}_y + \lambda_0\chih_2    &=& 0   \label{Ey}
\ear
where $\lambda_0$ is the separation constant.
Solutions to (\ref{Ey}), finite on the symmetry axis $y=\pm 1$,
are the Legendre polynomials $P_l (y)$, while
$\lambda_0=l(l+1)$ with $l=0,1,2,\ldots$. The corresponding solutions to
(\ref{Ex}) are combinations of Legendre functions of the first and
second kinds.

     The equations for $\nu$ and $W$ are solved in a similar way.

	This is the way to obtain a general class of solutions
	containing arbitrary multipolarities $l$: after writing out the
	solutions to the three Laplace equations (each has, in general, the
	form of an infinite series --- a superposition of different
	multipolarities), $\beta(\rho, z)$ is found by quadratures from the
	equations (obtained from (\ref{Edil}))
\bear                                                   \label{beta-xy}
	\beta_x  &=&  \frac{d}{2a_1}\Delta_x \chi
			   + \frac{1}{a_2}\Delta_x \nu + K_1\Delta_x 
W  	 \nn
	\beta_y  &=&  \frac{d}{2a_1}\Delta_y \chi
			   + \frac{1}{a_2}\Delta_y \nu + K_1\Delta_y W
\ear
	where, for any function $f$,
\bear                                                   \label{Delta-xy}
	\Delta_x f &=& \frac{1}{L^2(x^2 + y^2)}
	\Bigl[x\rho^2 f_x^2 -2y\rho^2 f_x f_y -L^2 x(1-y^2)^2 f_y^2 \Bigr],\nn
	\Delta_y f &=& \frac{1}{L^2(x^2 + y^2)}
	\Bigl[L^2y(x^2+1)^2 f_x^2 + 2x\rho^2 f_x f_y -y\rho^2 f_y^2 \Bigr].
\ear
	In what follows, however, we restrict ourselves to $l=0$ (monopole
	solutions), under the asymptotic flatness conditions:
	$\gamma=\xi=\varphi=U=0$ at spatial infinity.

\subsection{Monopole solutions}  

	A monopole solution to Eq.\,(\ref{Ey}),
	regular at $y = \pm 1$, is just a constant, so that $\chi=\chi(x)$.
	Eq.(\ref{Ex}) takes the form $(x^2+\varepsilon)d\chi/dx=\const$. Its
	integration leads to the following expressions for
	$\chi(x)$ satisfying the asymptotic flatness condition:
\beq                                       
\chi = \vars {-\half q_1 \ln{\dst\frac{x+1}{x-1}},
                           \quad &\varepsilon=-1,\\[2pt]
              -q_1/x,               &\varepsilon=0,\\[2pt]
              -q_1 \arccot x, \quad & \varepsilon =+1. }   \label{Chi}
\eeq
	The monopole solutions for $\nu$ and $W$ are found in a similar way
	and are described by (\ref{Chi}) with the replacements
\beq                                            
	q_1\ \mapsto \ q_2\quad {\rm for}\ \nu; \qquad
	q_1\ \mapsto \ q_3\quad {\rm for}\ W                     \label{QQ}
\eeq
	The expressions for $\beta(x,y)$ satisfying the asymptotic
	flatness condition $\beta(\infty,y)=0$ are
\bearr
     \e^{2\beta} = \vars {
     (x^2-1)^K (x^2-y^2)^{-K},\ &\varepsilon=-1,\\[3pt]
     \exp[-K(1-y^2)/x^2],   &\varepsilon=0,\\[3pt]
	(x^2+y^2)^K(x^2+1)^{-K}, & \varepsilon=+1  }
	                                               \label{Beta}   \\ &&
     K=\frac{d}{2a_1}q_1^2 + \frac{1}{a_2}q_2^2 + K_1 q_3^2,   \label{K}
\ear
     where $K_1$ is defined in (\ref{DefAK}).

     At spatial infinity our monopole solutions are asymptotically \sph.
     Indeed, assuming $y=\cos \theta$, where $\theta$ is the conventional
     polar angle, the line element (\ref{DsA}) transformed by (\ref{XY}) is
	\sph if
\beq                                                             
	\e^{2\beta} = 	(x^2 + \varepsilon)/(x^2 + \varepsilon y^2).
	                                                     \label{Sph}
\eeq
The condition (\ref{Sph}) holds for all the solutions in the limit
$x\to\infty$ where they have Schwarzschild asymptotics.
As for the whole space, the condition (\ref{Sph}) is fulfilled under
the additional requirement $K=-\varepsilon$. Unlike the vacuum case,
when $K$ is positive-definite and the above condition can hold for
$\varepsilon=0, +1$ only in the trivial case when the space-time is
flat, in the presence of an electric field $K$ can have any sign,
hence the sphericity condition can be fulfilled with any $\varepsilon$.
One naturally obtains the known solutions
described in \cite{br-vuz91,br95a}.

The solutions with $\varepsilon=0,\ -1$ turn out to possess naked
singularities in all nontrivial (nonspherical) cases.
So let us pay more attention to the solution with $\varepsilon=+1$,
which can have no curvature singularity.
Although a preferred conformal gauge does exist (the
string one), it is remarkable that the most important features of these
configuration are insensitive to conformal factors of the forms
$\exp(\const\times\sigma)$ and $\exp(\const\times\varphi)$, due to
global regularity of both $\varphi$ and $\sigma$.

\section{Ring wormholes}  
\subsection{The wormhole geometry}

	For $\varepsilon=+1$, the 4-dimensional scalars $\chi$, $\nu$ and
	$W$ are finite functions of $x$:
\beq                                                         \label{arccot}
	(\chi,\ \nu,\ W) = -(q_1,\ q_2,\ q_3)\arccot x.
\eeq
	Thus the metric $\go_{\mu\nu}$ and the field
	$F$ are regular if and only if $|\omega| < \infty$, i.e.,
     if $U$ nowhere tends to infinity or a root of the trinomial (\ref{Om}).
     If we require that at spatial infinity $U=\omega=0$, then $c_0=1/a_3$,
	$K_1= a_3 c_1^2 -1$ and the trinomial's discriminant is
	$4a_3^2(c_1^2-c_0) = 4a_3 K_1$, so that (\ref{Om}) gives:
\beq
	W-W_0 = \vars{
            -{\dst\frac{1}{2\sqrt{a_3K_1}}}
     \ln
{\dst\frac{U+c_1-\sqrt{K_1/a_3}}{U+c_1+\sqrt{K_1/a_3}}},\quad & K_1>0,\\
     -{\dst\frac{1}{a_3(U+c_1)}},\quad c_1^2=a_3,      & K_1=0,\\
     -{\dst \frac{1}{\sqrt{-a_3K_1}}}
    \arccot{\dst\frac{U+c_1}{\sqrt{-K_1/a_3}}}, & K_1<0,  }  \label{WU}
\eeq
where $W_0=\const$ must be chosen to satisfy $W=0$ for $U=0$, according
to (\ref{arccot}).

Evidently $U$ cannot tend to a root of (\ref{Om}), since
otherwise we would get $W\to\infty$, whereas actually
$W(x)$ ranges over a finite interval. To prevent $U\to \pm\infty$
it is sufficient to require that the corresponding value of
$W$ do not belong to this interval. We will now assume that this
condition holds and postpone its further discussion till Subsec.\,3.2.

The absence of a curvature singularity does not necessarily mean,
however, that the space-time is globally regular. Let us study the limit
$x\to 0$ in some detail.

The functions $\varphi,\ \xi,\ \gamma$ and $\e^\beta$ are finite at
$x=0$. The curve $x=0,\ y=0$ lies in
the plane $z=0$ and forms a ring $\rho=L$ of finite length (Fig.1).

\unitlength=1.2mm
\begin{figure}
\centering
\begin{picture}(83,50)
	\put(41.5,3){\vector(0,1){44}}
	\put(11.57,12){\circle*{2.5}}
	\put(71.5,12){\circle*{2.5}}
	\put(41.5,11){\circle*{1.2}}
	\put(41.5,13){\circle*{1.2}}
\thicklines
	\put(11.5,11){\line(1,0){60}}
	\put(11.5,13){\line(1,0){60}}
	\put(43,44){$z$}
	\put(9,6){$A$}
	\put(72,6){$B$}
	\put(9,16){\shortstack[l]{$x=0$\\ $y=0$\\ $\phi=\pi$}}
	\put(70,16){\shortstack[l]{$x=0$\\ $y=0$\\ $\phi=0$}}
	\put(42.5,16){\shortstack[l]{$x=0$\\ $y=+1$}}
	\put(42.5,4){\shortstack[l]{$x=0$\\ $y=-1$}}
\end{picture}
\caption{\protect\small
     Axial section of the neighborhood of the ring $x=y=0$.
     The points $A$ and $B$, marked by big black circles,
     belong to the ring, the thick lines connecting them show the upper
     and lower disks $x=0,\ y{> \atop <}0$. }
\medskip \hrule
\end{figure}
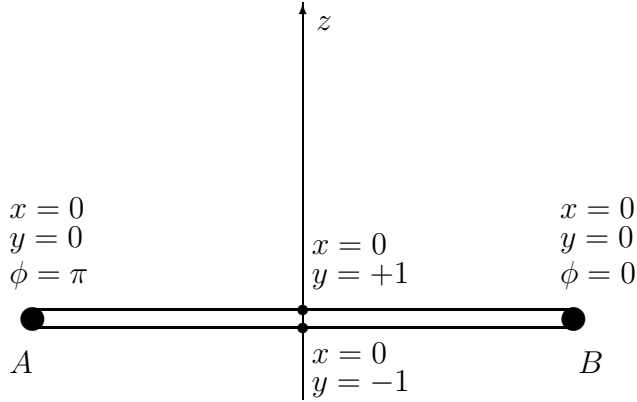

The surface $x=0,\ y>0$ is a disk bounded by the
above ring and parametrized by the coordinates $y$ and $\phi$.
This metric is flat if and only if $K=0$. Otherwise the disk is curved
but has a regular center at $y=1$ (the upper small black circle in
Fig.\,1). The limit $x\to 0$ corresponds to approaching the disk from
the half-space $z>0$.

Another similar disk, the lower half-space one, corresponds to $y<0$.
The two disks are naturally identified when our oblate
spheroidal coordinates are used in flat space. In our case,
a possible identification of points $(x=0,\ y=y_0,\ \phi=\phi_0)$
and $(x=0,\ y=-y_0,\ \phi=\phi_0)$, where $\phi_0$ is arbitrary and
$0 < y_0 \leq 1$, leads to a finite discontinuity of the
extrinsic curvature of the surfaces identified, or, physically, to a
finite discontinuity of forces acting on test particles. This may be
interpreted as a membrane-like matter distribution, bounded by the
ring $x=y=0$.

The field discontinuity across the surface $x=0$ is avoided if one
continues the coordinate map to negative $x$, as is done for the
vacuum case in Ref.\,\cite{br95b}.
As a result, appears another ``copy'' of the
3-space, so that a particle crossing the regular disk $x=0$
along a trajectory with fixed $y$, threads a path through the ring and
can ultimately get to another flat spatial infinity, $x\to -\infty$
with new asymptotic values of $\gamma$, $\xi$ and $\varphi$.  The
function $\beta$ is even with respect to $x$ and hence coincides at
both asymptotics. We obtain a wormhole configuration, nonsymmetric with
respect to its ``neck'' $x=0$, having no curvature discontinuity,
except maybe the ring $x=y=0$.

To study the geometry near the ring, let us
consider a 2-surface of fixed $\phi$ at	small $x$ and $y$. In the polar
coordinates $(r,\psi)$ ($x=r\cos\psi$, $y=r\sin\psi$), and still
new ones $\mu$ and $\eta$ defined by
\beq
     r=[(K+2)\mu]^{1/(K+2)},\quad \psi=\eta/(K+2),  \label{PolarT}
\eeq
	the corresponding 2-dimensional metric is (up to a constant factor)
\beq
	dl^2_{(x,y)} = r^{2K+2}(dr^2 + r^2 d\psi^2) = d\mu^2 + \mu^2 d\eta^2
							   \label{Polar}
\eeq
     Thus the metric near the ring is locally flat. However, it is locally
flat on the ring itself only if $\eta$ ranges over a segment of length
$2\pi$. Is this the case?

Given $x>0$, the polar angle $\psi$ is defined on the segment
$[-\pi/2,\ \pi/2]$, hence $\eta \in [-\pi -K\pi/2,\ \pi+K\pi/2]$.
Consequently, the ring enjoys local flatness
only in the simplest case $K=0$, when $\beta = \const$. If the
electric field were absent, that would mean that $q_1=q_2=0$ and
the space-time is flat \cite{br95b}; however, with $q_3\ne 0$, $K=0$ no
longer implies	the global flatness (see (\ref{K})).

	In the wormhole case $x$ can have either sign, hence
\beq
	\psi\in [-\pi,\ \pi]\
	         \Rightarrow \eta\in [-(2+K)\pi,\ (2+K)\pi]  \label{Eta}
\eeq
Thus the axially symmetric wormhole solution contains in general a
cosmic string-like ring singularity with a polar angle excess greater or
smaller than $2\pi$ for $K>0$ and $K<0$, respectively. The case $K=0$
actually means that there is no singularity and the ring geometry
exactly corresponds to what should occur near the neck of such a
wormhole. Indeed, making one revolution through the ring
at fixed $\phi$ (completing an angle $\eta_2-\eta_1=2\pi$), an
observer finds himself at a similar position, but in the ``second
world'', with (in general) other values of $\varphi$, $\xi$ and $W$,
and returns to his original position only completing an angle of $4\pi$.

For $K=0$ the $(x,y)$ surface behaves near $x=y=0$ like the
two-sheeted Riemann surface of the analytic function $\sqrt{x+iy}$;
and, indeed, the above transition $(x,y)\equiv (r,\psi) \to (\mu,
\eta)$ may be described as a conformal mapping in the complex plane
with this analytic function: $r\e^{i\psi} = \sqrt{\mu\e^{i\eta}}$.
Other examples of branching points in the space-time are known in some
Einstein-Maxwell fields in 4 dimensions \cite{br79}.

\subsection{Wormhole parameters}

Let us use for further interpretation the string metric (\ref{MapS}),
where the 4-dimensional part $g_{\mu\nu}$ of $g_{AB}$ is connected with
our $\go_{\mu\nu}$ by
\beq
	\go_{\mu\nu} = \e^{d\cdot \xi} g_{\mu\nu}.  \label{MapE}
\eeq
	(note that conformal mappings do not affect angles, so that
	the ring regularity condition does not suffer).
	Consequently, the string metric is
\beq
	\ds = \e^{2\hx} ds_1^2 + \e^{2\omega}dt^2
            - \e^{2\omega -4\gamma}\bigl[ e^{2\beta}(d\rho^2 + dz^2)
              + \rho^2 d\phi^2\bigr]               \label{DsS}
\eeq
	where
\beq
	\hx = \xi - \lambda \varphi =
          -\frac{d+2}{a_1}\lambda\chih+\frac{2}{a_1 a_2}
          [1-(d+2)\lambda^2] (\nu - \omega)        \label{XiS}
\eeq
	and the other functions are defined as before.  Note that for
	$\lambda=\lst$ one has just $\hx = - \chih$.

	Our solution depends on four integration constants (the ``charges''
	$q_i$ and $c_1$), two input parameters $d$ and $\lambda$ and the
	length scale  $L$ which appeared in the transformation  (\ref{XY})
	and, like the integration constants, is a parameter of the family of
	solutions.

	Let us connect the wormhole electric charge and mass and their other
	properties with the input and integration constants.

     First, let us restrict ourselves to regular wormholes, for which $K=0$,
     hence $K_1\leq 0$ ($K_1 =0$ is suitable only in the case $q_1= q_2=0,\
     q_3\ne 0$). The case of all $q_i=0$ is trivial and reduces to flat
     space without fields.

Second, we will try to select symmetric wormholes, i.e., those with
equal asymptotic values of the time rate, i.e.,
$\e^{\omega(\pm\infty)} =1$; other symmetry
relations will then follow. An equal time rate means, in particular, a
possibility to ``immerse''  the two asymptotics (``mouths'') of a
wormhole to the same external Minkowski space-time without clock
rescaling.

	Let us fix
\beq
	-K_1= 1-a_3 c_1^2 = \cos^2 \mu >0,\qquad
	\sin\mu = \sqrt{a_3} c_1                            \label{Mu}
\eeq
	where $\mu$ is a new parameter introduced for convenience. By
	(\ref{Om}), $\e^{2\omega}$ is positive for all $U$ and, as $W(x)$ 
is a
	monotone function and $dW/dU=\e^{-2\omega}$, $U(x)$ is also a monotone
	function. This immediately implies that {\sl the electric charges of
	the wormhole as seen from the two asymptotics} (determined, up to
	a positive factor, by $dU/d|x|$) {\sl have different signs}.

	The third line of (\ref{WU}) can be rewritten as
\beq
	\sqat\frac{U+c_1}{\cos\mu}=
	- \tan\Bigl[ \sqat\cos\mu (q_3\arccot x + W_0)\Bigr]  \label{UW}
\eeq
	where $W_0$ is a constant determined from the condition
	that $U$ and $W$ both vanish as $x\to\infty$:
\beq
	\tan\mu = -\tan\bigl[\sqat\cos\mu\cdot W_0\bigr] \quad
	\Rightarrow  \quad W_0 = -\mu\Big/ (\sqat\cos\mu).          \label{W0}
\eeq

 	The expressions for masses and charges can be obtained from the field
	behavior at the asymptotics. Thus, for $x\to\infty$ the mass 
$M_+$ is
	determined by the Schwarzschild asymptotic of $\g_{\mu\nu}$:
\beq
	\e^{2\omega} = 1 - 2GM/r_+ + o(1/r_+)        \label{Mass}
\eeq
 	where $r_+$ is the asymptotic expression for the spherical radius:
 	$r_+ = Lx$. Comparing (\ref{Mass}) with the actual asymptotic of
	$\e^{2\omega}$, one obtains
\beq
	GM_+ = Lc_1a_3q_3.                          \label{Mplus}
\eeq
  	Thus $M_+$ is positive if and only if the constants $c_1$ and $q_3$
  	(and hence also $\mu$) have the same sign.

The charge $Q_+$ at the same asymptotic is determined by the electic
field behavior:	the asymptotic field magnitude must be
\beq
	E^i = \sqrt G Q_+ r_+^i/r_+^3    \label{xvii}
\eeq
	where the index $i$ enumerates 3-vector components in the
     asymptotic Minkowski space and the appearance of $\sqrt G$ conforms to
     the way the electromagnetic field was introduced in the Lagrangian. A
     comparison to the actual field shows that
\beq
	\sqrt G Q_+ = Lq_3.        \label{xviii}
\eeq

     Comparing (\ref{Mplus}) and (\ref{xviii}), one obtains a necessary
	condition for wormhole regularity in terms of masses and charges:
\beq
	GM_+/Q_+^2 = a_3 \sin^2 \mu
	        < a_3 = 1 + \frac{d}{d+2} + 2\lambda^2.   \label{xix}
\eeq
	The same condition could be obtained on equal grounds for the
     asymptotic $x\to -\infty$ but with slight complications connected with
     the other appearance of the boundary conditions.

Another restriction on the solution parameters is connected with the
dependence $U(x)$ (\ref{UW}),(\ref{W0}). By construction, the
argument of the tangent in the r.h.s. of (\ref{UW}) is within the
interval $(-\pi/2,\ \pi/2)$ at $x\to\infty$ (it just equals $\mu$).
For $U(x)$ to remain regular at all $x$, the above argument should not
ever leave this interval. Taking into account that, for $M_+ > 0$,
the quantities $\mu$, $c_1$ and $q_3$ have the same sign, it is easy to
obtain this second regularity condition in the form
\beq
	\pi\sqat\cos\mu \cdot |q_3| - |\mu| < \pi/2,     \label{xx}
\eeq
	or
\beq
	|q_3| < (\pi/2 + |\mu|)\Big/ (\pi\sqat\cos\mu),  \label{xxi}
\eeq
a restriction upon $q_3$ provided the parameter  $\mu$ (connected with
the charge-to-mass ratio) is specified.

The combined conditions $K=0$ (whence follows, in particular,  (\ref{xix})),
and (\ref{xx}) or (\ref{xxi}) are not only necessary, but also sufficient
for our solution's regularity.

     We did not so far apply the symmetry condition. Putting in (\ref{Om})
     $\e^{2\omega}=1$ and recalling that $a_3c_0 =1$, we find two possible
	values for $U$:  $0$ and $-2c_1$.  As $U=0$ corresponds to $x\to
	+\infty$, for a nontrivial solution $U(-\infty) = -2c_1$. Let us
	substitute this into (\ref{UW}), then, with (\ref{W0}),
\beq
	\pi\sqat q_3 \cos\mu = 2\mu.                  \label{xxii}
\eeq
	Eq.\,(\ref{xxii}) has a number of consequences:
\begin{description}
\item{(i)}
	(\ref{xxii})  implies that $\mu$ and $q_3$ have the same sign.
     That is, {\sl if a wormhole is regular and symmetric, it can have only
     a positive mass\/} (and it is positive as seen from either of the
     asymptotics).
\item{(ii)}
	Given (\ref{xxii}), (\ref{xx}) is fulfilled automatically. Thus
     {\sl $K=0$ plus symmetry is sufficient for regularity.}
\item{(iii)}
	(\ref{xxii}) connects $q_3= \sqrt G Q_+/L$ with $\mu$, or with the
     charge-to-mass ratio. In particular, if we know the mass and the charge,
     (\ref{xxii}) yields the relevant length scale $L$ of the solution:
\beq
	L = \pi\sqrt{a_3 G} Q_+ \cos\mu/(2\mu).            \label{xxiii}
\eeq
	For big charge-to-mass ratios, such as those known for particles
     (about $10^{18}$ for a proton), the gravitational constant in
     (\ref{xxiii}) is cancelled and we obtain $L$ of the order of the
     ``classical radius'' $Q^2/M$:
\beq                                                    \label{clas}
	L \approx \pi a_3 Q_+^2/M_+,
\eeq
	which may be treated as a hint that
     regular soliton-like particle models can be obtained on this trend.
\end{description}

With (\ref{xxii}), the quantity $U+c_1$ becomes an odd function of $x$,
\beq
	U+c_1 = \frac{\cos\mu}{\sqat}
	\tan\Bigl(\frac{2\mu}{\pi}\arctan x\Bigr).
\label{xxiv}
\eeq
	The function $\omega(x)$ is even:
\beq
	\e^{2\omega} = a_3(U+c_1)^2 + \cos^2 \mu.          \label{xxv}
\eeq
However, the whole solution is not entirely symmetric with respect to $x=0$.
Indeed, since at least one of the charges $q_1$ and $q_2$ is nonzero,
there is a nontrivial monotone $x$-dependence of $\chih$ or $\nu$ (or both).
Consequently, either the $\varphi$ field, or the internal space scale
factor $\hx$, or both have different finite limits at $x=+\infty$ and
$x=-\infty$. By the conventional terminology, the solution behaves as a
kink with respect to these fields. (An attempt to find a relation
between $d$ and $\lambda$ such that the $x$-dependence of $\varphi$ and
$\hx$, other than via $\omega(x)$, would disappear, fails since the
emegring set of algebraic equations has no solution.) As is easily verified,
this nonsymmetry conclusion applies to all conformal gauges connected by
functions of $\varphi$.

Finally, let us address to the ``purely electric wormhole'' solution
$q_1 = q_2 =0,\ q_3\ne 0,\ K=K_1 =0$, also satisfying the regularity
condition connected with the geometry near the ring $x=y=0$.
In this case all fields are expressed  in terms of $\omega(x)$:
\beq
    \varphi= -2\lambda\omega/a_3, \qquad
    \xi = -2\omega/(a_1 a_2), \qquad
    \gamma = \omega/a_3.                \label{xxvi}
\eeq
	The condition $K_1=0$ implies $c_1= \pm 1/\sqat$. Eq.\,(\ref{WU})
     subject to the condition that $U$ and $W$ both vanish as $x\to\infty$
     yields
\beq
	U= c_1 W/(c_1-W),               \label{xxvii}
\eeq
     where $W$ is, as before, expressed by (\ref{arccot}).
     The regularity condition for $U$ (that $W\ne c_1$ for any real $x$) is
     valid if
\bear
     -c_1/q_3 <0  \qquad && (\rm for\ \sign c_1 = \sign q_3), \nn
     -c_1/q_3 > \pi \qquad && (\rm for\ \sign c_1 \ne sign q_3).
            \label{xxviii}
\ear
	Thus regular solutions are obtained. However, the explicit form
     of $\omega(x)$
\beq
     \e^{2\omega} = \Bigl(1+\frac{U}{c_1}\Bigr)^2
      = \Bigl(1 + \frac{q_3}{c_1} \arccot x\Bigr)^{-2}   \label{xxix}
\eeq
	shows that in the nontrivial case $q_3\ne 0$ the function
     $\omega(x)$ is monotone. That means that not only the wormhole 
cannot be
	symmetric, but the mass has different signs at the two asymptotics.
	Hence this solution can hardly be treated as a satisfactory one,
	despite its regularity.

\section{Concluding remarks}

{\bf 4.1.}
The ring wormholes described here are a type of nonsingular
Lorentzian wormholes able to exist in spaces of any dimension.
A somewhat similar solution --- stationary, \axi, 5-dimensional,
was found by G. Clement in Ref.\,\cite{cle1}, but there the function
$\e^\beta$ (in our notation) grew to infinity as $x\to 0,\ y\to 0$ and
it was concluded that ``the singular ring is at infinity''. The author
mentioned three problems with the physical interpretation of this
solution. First, the solution parameters, including the mass, as
measured at $x\to -\infty$, are the opposite of those measured at
$x\to \infty$ (though, the isometry exchanging the ``end points" might
be interpreted as classical mass and charge conjugation).  Second,
there was no definite sign of the mass. And third, the existence of
unobserved long-range scalar forces.

For the present symmetric \wh s the first two problems do not exist:
the mass is manifestly positive ``from both sides". The third problem
is common for all models with long-range scalar fields, be they
of dilatonic or extra-dimensional origin, or Goldstone, or Brans-Dicke
ones, etc. We would like just to recall that the existence of at least
one such field ($\varphi$ or $\xi$) in our model is necessary in order
to provide the \wh\ regularity, in other words, to maintain a kind of
equilibrium with the electromagnetic field.

\medskip\noindent
{\bf 4.2.} Unlike the \sph case, where \wh s
can appear only with matter violating the usual energy conditions,
e.g., unconventional scalar fields (\cite{br-acta,hell,cle}  and
others) or special kinds of nonlinear electrodynamics \cite{bryba},
under axial symmetry they appear rather	naturally in vacuum,
scalar-vacuum and electrovacuum systems both in GR and
dilaton gravity. Other types of nonspherical \wh s, connected with
cosmic strings, are considered in Ref.\,\cite{clem}.

\medskip\noindent
{\bf 4.3.} A tentative study shows that the present class of regular \wh\
models can be generalized to include the harmonic functions (``scalar
fields") $\nu,\ \chi,\ \omega$  with higher multipole moments
($l > 0$), which may be treated as perturbations to purely monopole
models. The higher-$l$ parts of the fields are strongly constrained by
the asymptotic flatness conditions and those of regularity at
the ring $x=y=0$. We hope to give a more detailed treatment of such
models elsewhere.

\medskip\noindent
{\bf 4.4.} For the present class of static, \axi configurations,
addition of new scalar fields to electrically neutral models (for
instance, to pure vacuum in GR) affects the properties of the
solutions only quantitatively; this applies, in particular, to
extra-dimension scale factors which, in the 4-dimensional setting
of the problem, actually behave as massless,
minimally coupled scalar fields.  Unlike that, the electric field (all
the same in GR or dilaton gravity) creates qualitatively new features:
there appear new types of singularities and globally regular \axi
wormholes.  See, however, the last sentence of item 4.1.

\medskip\noindent
{\bf 4.5.} The 4-dimensional version of our models with $\lambda=0$ are
readily reformulated in terms of a broad (Bergmann-Wagoner) class of
scalar-tensor theories (STT) of gravity in 4 dimensions, in particular,
the Brans-Dicke theory and GR with a conformally coupled scalar field.
Models of different STT differ from each other by conformal mappings
with regular ($\varphi$-dependent) conformal factors which change the
explicit form of \wh\ symmetry conditions but do not affect the neck
regularity. Note that the scalar field of the STT can be interpreted in
terms of an extra-dimension scale factor, while different theories
correspond to different conformal gauges of the 4-dimension section.
In the case $\lambda\ne 0$ one can speak of generalized STT with a
coupling between the scalar and electromagnetic fields.

To conclude, both globally regular solutions and those with ``strings"
may be of interest for describing late stages of gravitational
collapse and/or cosmological dark matter. Their monopole nature
probably means that	they cannot decay by gravitational wave emission.
So it can be of significant interest to study their generalizations and
stability.

\subsection*{Acknowledgement}

This work was supported in part by the Russian Ministry of Science
and by CNPq Grant No. . One of us (K.B.) is sincerely grateful to the
Department of Physics of UFES for kind hospitality.

\end{document}